# A massive proto-cluster of galaxies at a redshift of z ≈ 5.3

Peter L. Capak[1], Dominik Riechers[2,3], Nick Z. Scoville[2], Chris Carilli[4], Pierre Cox[5], Roberto Neri[5], Brant Robertson[2,3], Mara Salvato[6], Eva Schinnerer[7], Lin Yan[1], Grant W. Wilson[8], Min Yun[8], Francesca Civano[9], Martin Elvis[9], Alexander Karim[7], Bahram Mobasher[10], & Johannes G. Staguhn[11]

[1]Spitzer Science Centre, 314-6 California Institute of Technology, 1200 E. California Blvd., Pasadena, CA, 91125, USA (capak@astro.caltech.edu)
[2]Department of Astronomy, 249-17 California Institute of Technology, 1200 E. California Blvd., Pasadena, CA, 91125, USA
[3]Hubble Fellow
[4]National Radio Astronomy Observatory, PO Box O, Socorro, NM, 87801, USA
[5]Institut de Radio Astronomie Millimétrique, 300 rue de la Piscine, F-38406 St-Martin-d'Hères, France
[6]Max-Planck-Institute für Plasma Physics, Boltzmann Strasse 2, Garching 85748, Germany
[7]Max-Planck-Institute für Astronomie, Königstuhl 17, Heidelberg 69117, Germany
[8]Department of Astronomy, University of Massachusetts, Lederle Graduate Research Tower B, 619E, 710 North Pleasant Street, Amherst, MA, 01003-9305, USA
[9]Harvard Smithsonian Center for Astrophysics, 60 Garden Street, MS, 67, Cambridge, MA, 02138, USA
[10]Department of Physics and Astronomy, University of California, Riverside, CA, 92521, USA
[11]Johns Hopkins University, Laboratory for Observational Cosmology, Code 665, Building 34, NASA's Goddard Space Flight Center, Greenbelt, MD, 20771, USA

**Massive clusters of galaxies have been found as early as 3.9 Billion years (z=1.62)[1] after the Big Bang containing stars that formed at even earlier epochs[2,3]. Cosmological simulations using the current cold dark matter paradigm predict these systems should descend from "proto-clusters" – early over-densities of massive galaxies that merge hierarchically to form a cluster[4,5]. These proto-cluster regions themselves are built-up hierarchically and so are expected to contain extremely massive galaxies which can be observed as luminous quasars and starbursts[4,5,6]. However, observational evidence for this scenario is sparse due to the fact that high-redshift proto-clusters are rare and difficult to observe[6,7]. Here we report a proto-cluster region 1 billion years (z=5.3) after the Big Bang. This cluster of massive galaxies extends over >13 Mega-parsecs, contains a luminous quasar as well as a system rich in molecular gas[8]. These massive galaxies place a lower limit of >4x10$^{11}$ solar masses of dark and luminous matter in this region consistent with that expected from cosmological simulations for the earliest galaxy clusters[4,5,7].**

Cosmological simulations predict that the progenitors of present day galaxy clusters to be the largest structures at high redshift ($M_{halo}$> 2x10$^{11}$ solar masses ($M_\odot$), $M_{stars}$ > 4x10$^9$ $M_\odot$ at z~6)[4,5,7]. These proto-cluster regions should be characterized by local over-densities of massive galaxies on 2-8 co-moving Mega-parsecs (Mpc) scales that coherently extend over 10s of Mpc, forming a structure that will eventually coalesce into a cluster[4,5,7,9]. Furthermore, due to the high mass densities and correspondingly high merger rates, extreme phenomena such as starbursts and quasars should preferentially exist in these regions[4,5,6,7,9,10]. Although over-densities have been reported around radio galaxies on ~10-20 Mpc[6,7] scales and large gas masses around quasars[11,12] above z=5, the data at hand is not comprehensive enough to constrain the mass of these proto-clusters and hence provide robust constraints on cosmological models[6,7,9].

We used data covering the entire accessible electromagnetic spectrum in the 2 square degree COSMOS field (RA=10:00:30, DEC=2:30:00)[13] to search for starbursts, quasars, and massive galaxies as signposts of potential over-densities at high redshift. This deep, large area field provides the multi-wavelength data required to find proto-clusters on >10 Mpc (5 arc minute) scales. Optically bright objects above a redshift of 4 were identified through optical and near-infrared colours. Extreme star formation activity was found using millimetre-wave[14,15] and radio[16] measurements, and potential luminous quasars were identified by X-ray measurements[17]. Finally, extreme objects and their surrounding galaxies were targeted with the Keck-II telescope and the Deep Extragalactic Imaging Multi-Object Spectrograph (DEIMOS) to measure redshifts.

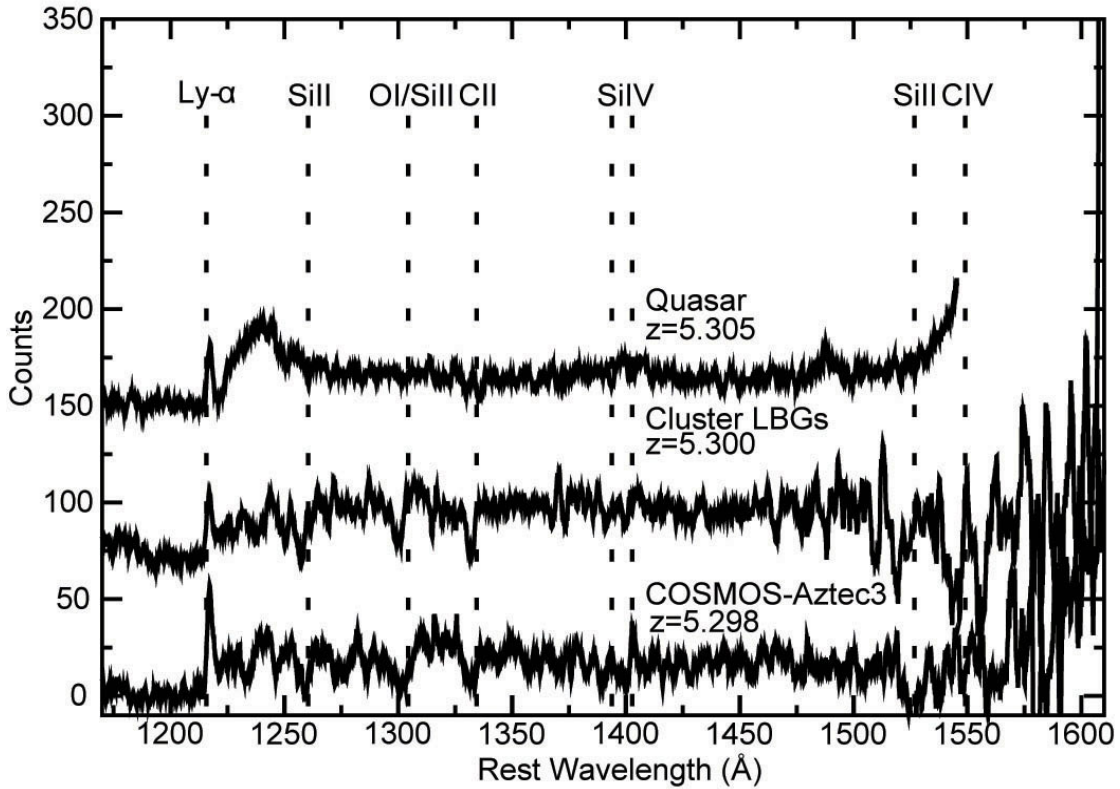

**Figure 1: Spectra of confirmed cluster members.** These spectra were taken with the Keck-II telescope and correspond to the extreme starburst (COSMOS-AzTEC3), a combined spectra of two Lyman break galaxies at 95kpc (Cluster LBGs), and the Chandra detected quasar at 13Mpc from the extreme starburst. The galaxy spectra show absorption features indicative of interstellar gas (SiII, OI/SiII, CII) and young massive stars (SiIV, C[IV]) indicative of a stellar population less than 30Myr old[26]. The quasar shows broad Ly-$\alpha$ emission absorbed by strong winds with a narrow Ly-$\alpha$ line seen at the same systemic velocity as absorption features in the spectra.

We found a grouping of four major objects at z=5.30 shown in Figure 1. The most significant over-density appears near the extreme starbursting galaxy COSMOS AzTEC-3 which contains >$5.3 \times 10^{10}$ $M_\odot$ of molecular gas and has a dynamical mass, including dark matter, of >$1.4 \times 10^{11}$ $M_\odot$[8]. The far infrared luminosity (60-120μm) of this system is estimated to be 1.7±0.8 x$10^{13}$ solar luminosities (L$_\odot$) corresponding to a star formation rate of >1500 M$_\odot$ per year[18], >100 times the rate of an average (L$_*$) galaxy at z=5.3[19]. The value and error given is the mean estimate and scatter derived from empirical estimates based on the sub-mm flux, radio flux limit, and CO luminosity, along with model fitting. The models predict a much broader range in total infrared luminosities (8-1000µm) ranging between 2.2-11x$10^{13}$ L$_\odot$. The large uncertainty results from the many assumptions used in the models combined with a lack of data constraining the infrared emission blue-ward of rest-frame 140µm. However, the observed limit on the sub-millimeter spectral slope favours models with colder dust, and hence lower luminosities.

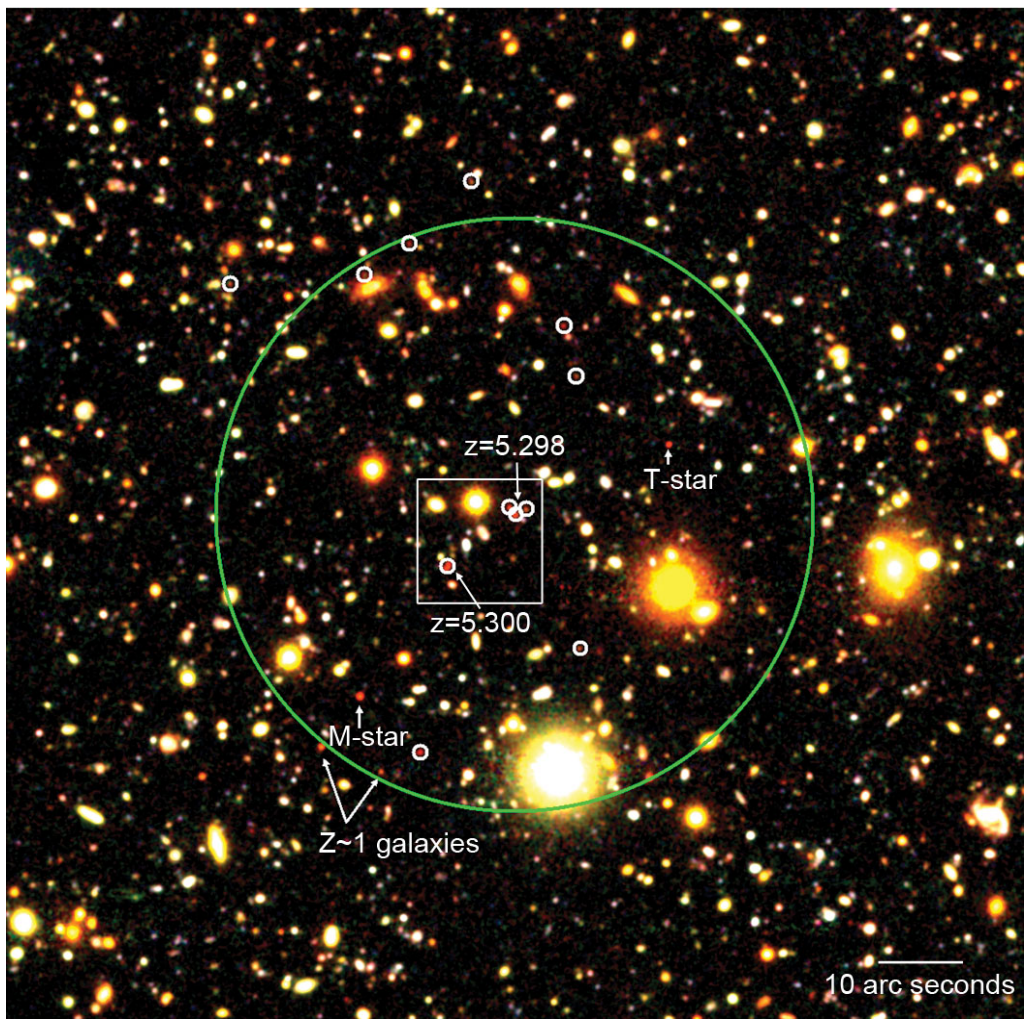

**Figure 2: Image of the region around the proto-cluster core.** This area corresponds to a 2'x2' region around the starburst (COSMOS AzTEC-3). The z~5.3 candidates are marked in white and a 2Mpc co-moving radius is marked with a green circle. The area highlighted in Figure 3 is denoted with a white box and the optical counterpart of the sub-mm source COSMOS AzTEC-3 is labeled "starburst". Spectroscopic redshifts and other red objects that have been identified as galactic stars or low redshift galaxies by their spectral energy distribution are also indicated with arrows.

The significance of the over-density around the starburst is immediately apparent in Figures 2 and 3. In the 1 square arc-minute (2.3 x 2.3 Mpc at z=5.3) around the starburst one would expect to find 0.75±0.04 bright ($z_{850}$<26) galaxies with colours consistent with a Lyman Break in their spectra at z=5.3[19], instead we find 8. This is a factor of 11 over-density, assuming the redshift range of 4.5<z<6.5 probed by typical broad band colour selections[19,20].

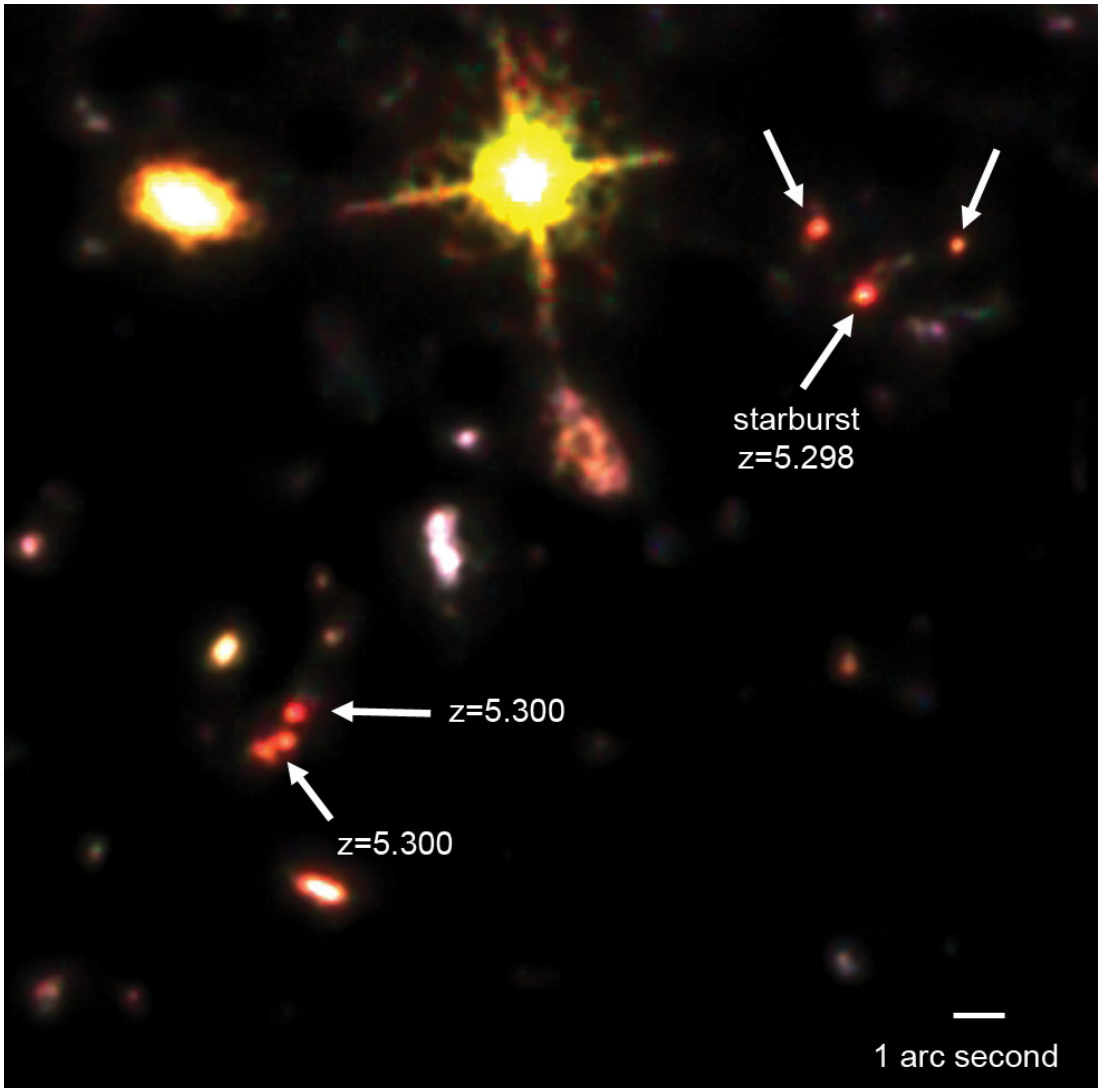

**Figure 3: An enlarged image of the proto-cluster core.** This area corresponds to 22.5" x 22.5" on the sky, 0.865 Mpc co-moving, or 0.137 Mpc proper distance at z=5.298. The six optically bright objects with spectral energy distributions consistent with z=5.298 are marked and spectroscopic redshifts are indicated. The optical counterpart of the sub-mm source COSMOS AzTEC-3 is labeled "starburst".

Within a 2 Mpc radius of the starburst we find 11 objects brighter than L* whose intermediate band colours[21] are consistent with being at z=5.3. This represents a >11 times over-density in both the measured and expected density of luminous galaxies. Estimates of the typical variance from clustering and cosmological simulations suggest this is significant at the >9σ level even if we only consider the spectrocopically confirmed systems. Of these 11 objects, three (including the optical counterpart of COSMOS AzTEC-3) are within 12.2 kpc proper distance (2") of COSMOS AzTEC-3, and two additional spectroscopically confirmed sources are found 95 kpc (15.5") away.

X-ray (0.5-10keV band) selected z>5 quasars are extremely rare[22] due to the high luminosities required for detection, yet one is found[17] within 13 Mpc of the starburst at the same spectroscopic redshift as COSMOS AzTEC-3. The distance between these objects is comparable to the co-moving distance scale expected for proto-clusters at z~5[5,7]. The optical spectrum of this object exhibits deep blue shifted gas absorption features indicative of strong winds driven by the energy dissipated from the rapid black hole growth. The object has a X-ray luminosity of $1.9 \times 10^{11}$ $L_\odot$ and a bolometric luminosity estimated from its spectral energy distribution of $\geq 8.3 \times 10^{11}$ $L_\odot$ (Hao et al. in preparation), implying a black hole mass of $\geq 3 \times 10^7$ $M_\odot$ if it is accreting at the Eddington rate, with a more likely mass of $\approx 3 \times 10^8$ $M_\odot$ for the typical accretion rate of 0.1 Eddington[5]. Assuming the final black hole to stellar mass relation of $M_{BH} \approx 0.002$ $M_\odot$[5] implies this object will eventually have a stellar mass $>10^{10-11}$ $M_\odot$, placing it amongst the most luminous and massive objects at this redshift[19,23].

The stellar mass of the proto-cluster system was estimated by fitting stellar population models to the rest-frame ultraviolet to optical photometry of the individual galaxies in the proto-cluster. The redshift was fixed at z=5.298, and models with a single recent burst of star formation[24] were used allowing for up to 10 visual magnitudes of extinction[25]. [OII] and Hα emission lines were added to the templates with fluxes proportional to the ultraviolet continuum of the template[18]. The accuracy of the stellar mass estimate is limited by the sensitivity of the 0.9-2.5μm photometry. The present data are insufficient to fully break the degeneracy between stellar age and dust obscuration. However, the age of 10 Myr derived from the photometric fitting is consistent with the features seen in the Keck spectra[26]. Given the range of acceptable fits and the concordance with the Keck spectra, the resulting stellar mass is probably accurate to a factor of ~2 (0.3 dex).

Using the described procedure we conservatively estimate that the starburst AzTEC-3 has a stellar mass of $1-2 \times 10^{10}$ $M_\odot$, implying the baryonic matter is >70% gas, nearly twice that found in typical star-bursting systems[27], but in agreement with the dynamical estimates[8,28]. The stellar masses of the 11 objects in the proto-cluster core contain a total of $>2 \times 10^{10}$ $M_\odot$ of stars, with individual galaxies weighing between $0.06-10 \times 10^9$ $M_\odot$. With this stellar mass and gas fraction a lower limit can be placed on the total mass of this system, assuming a global dark matter to baryon ratio of 5.9[1]. The resulting total halo mass is $>4 \times 10^{11}$ $M_\odot$, with the starburst residing in a halo of mass $>10^{11}$ $M_\odot$ comparable to the halo masses predicted for galaxies that will eventually merge into modern day galaxy clusters[7]. However, we note that the actual mass is probably much higher since much of the baryonic mass is likely in un-observed hydrogen gas, and the starburst object alone accounts for >37% of the total mass. Furthermore, the contribution of significantly more numerous fainter (L<L*) galaxies[19] are not counted in this mass estimate.

The three objects around COSMOS AzTEC-3 likely represent the progenitor of a massive central cluster galaxy (cD) at lower redshift. These objects are already within the radius of a typical local cD galaxy and their dynamical timescale is ~60 Myr, assuming a velocity dispersion of 200 km/s. Even considering the objects at 95kpc the dynamical timescale is less than 0.5Gyr, providing several dynamical times for a merger to occur by z~2 (i.e. 2 Gyr later). However, the observed stellar mass in these galaxies is significantly less than the ~$10^{11-12}$ $M_\odot$ in a typical local cD galaxy[7], indicating the majority of the stars have yet to form.

The properties of this proto-cluster are in qualitative and quantitative agreement with galaxy formation simulations[4,5]. The spatial extent, star formation rate per unit mass, and gas properties of the core structure around the extreme starburst are all similar to the predictions for massive galaxy formation in simulations. Furthermore, the properties of the quasar are also in agreement with the models for the later phases of massive galaxy formation when the quasar becomes visible. Finally, unlike previously described over-densities at z>5[6] we have strong spectroscopic and photometric evidence for a range of objects including massive, heavily star forming, and active galaxies. These are found both in the core of the structure and over a much larger area, indicating one can study the effects of environment on galaxy formation as early as z~5. We conclude that this region contains a large-scale baryonic over-density in the very early universe that will evolve into a high mass cluster like those observed at lower redshifts.

**Acknowledgements** These results are based on observations with: the W.M. Keck Observatory, the IRAM Plateau de Bure Interferometer, the IRAM 30m telescope with the GISMO 2mm camera, the Chandra X-ray observatory, the Subaru Telescope, the Hubble Space Telescope, the Canada-France-Hawaii Telescope with WIRCam and MegaPrime, the United Kingdom Infrared Telescope, the Spitzer Space Telescope, the Smithsonian Sub-millimeter Array Telescope, the James-Clerk-Maxwell Telescope with the AzTEC 1.1mm camera, and the National Radio Astronomy Observatory's Very Large Array. D.R. and B.R. acknowledges support from NASA through Hubble Fellowship grants awarded by the Space Telescope Science Institute. P.L.C. and N.Z.S. acknowledge grant support from NASA. G.W.W., M.Y., and J.S. acknowledge grant support of the NSF.

**Author Contributions** P.L.C. led the spectroscopic effort, reduced the spectroscopic and photometric data, and lead the scientific analysis including the optical and radio/mm fitting analysis and cluster properties. N.Z.S. lead the spectroscopic and photometric follow-up efforts. D.R., C.C., P.C. and R.N. assisted with the physical interpretation of the radio data. B.R. provided cosmological simulations to check the significance of the proto-cluster and the likelihood of finding it. M.S., L.Y. M.E., F.C. and B.M. carried out the Keck observations and assisted with the data reduction. E.S. reduced and analyzed the radio data. G.W. and M.Y. assisted with the sub-mm data analysis. F.C. and M.E. assisted with the X-ray data analysis. A.K. coordinated the 2mm observations. J.S. conducted and reduced the 2mm observations.

The authors have no competing financial interests.

Reprints and permissions information is available at www.nature.com/reprints/
Correspondence and requests for materials should be addressed to Peter L. Capak (e-mail: capak@astro.caltech.edu).

**Supplementary Information**

**1. IRAC Photometry**

Spitzer IRAC fluxes were measured from a combination of programs 20070 (COSMOS), 40801, 50310 and 61043 (SEDS) in the COSMOS field. The data for program 61043 were obtained during the Spitzer warm mission and only the data obtained in January 2010 were used for this paper. The average exposure time at the position of the proto-cluster is 6.3 hours per pixel.

The data reduction procedure closely follows that used for the COSMOS data (program 20070)[29]. We began our reduction using the corrected basic calibrated data (cBCD) provided by the Spitzer Science Center. The cBCDs were background matched using the contributed afrl_bcd_overlap IDL procedure, and then combined into a mosaic using the MOPEX software package. Cosmic rays, asteroids, and other transient artefacts were removed using the MOPEX temporal and box outlier modules. The images were combined using a drizzle interpolation with a PIXFRAC of 0.65 and an exposure time weighted co-addition. The resulting image has a 1 sigma, background limited, point source sensitivity of 38 nJy and 44 nJy in 3.6μm and 4.5μm respectively for the region around the proto-cluster.

Fluxes were measured by fitting an empirically measured point spread function (PSF) at the source position. Compact sources within 5 arc-minutes of the proto-cluster were used to construct the PSF model. For un-confused sources the PSF fit fluxes agreed well with 3.8" diameter aperture photometry measurements corrected to total flux via a statistical aperture correction. The new IRAC fluxes for the sources are given in Supplementary Table 1.

**2. 2mm Observations of COSMOS AzTEC-3**

The Goddard-IRAM Superconducting 2-Millimeter Observer (GISMO) is a bolometer camera optimized for 2-millimeter observations at the IRAM 30m telescope. GISMO uses a closed packed 128 element transition edge sensor (TES) detector array with an instantaneous field of view of 2'x4'. The observations of AzTEC-3 were obtained on April 12, 2010 with 100 minutes of integration. Flux calibrations were obtained to 18% accuracy with observations of Mars and pointing was verified by frequent observations of the nearby quasar J0948+003. The data were reduced and calibrated using the CRUSH-2 data reduction package.

We detect a source with 3.1σ significance based on the 18" beam-smoothed flux maps and the noise maps within the expected positional error of AzTEC-3. Including the statistical, calibration, and positional errors the detection of AzTEC-3 has a 99% statistical probability. We estimate the 2mm flux to be 3.7±1.4 mJy including uncertainty from the flux calibration. We note the 2mm flux is not corrected for possible flux boosting[14] since this correction factor requires knowledge of the 2mm number counts and these are not well measured.

## 3. Selecting Cluster Members

One of the great strengths of the COSMOS data are the extremely accurate (dz/(1+z)<0.01) photometric redshfits at z<1.5[30]. However, the techniques used to construct the photometric redshifts suffer from catastrophic failures when applied to high redshift and extreme sources such as those in our proto-cluster. Supplementary Figure 1 shows a comparison of the spectroscopic[31,32] and photometric[30] redshifts for objects brighter than i<26 including a sample of 168 spectroscopic redshifts at z>3 in the COSMOS field. We found >40% of luminous z>4 objects are placed at z<2 by the photometric redshift method, including the extreme starburst presented here. This is primarily due to photometric contamination from foreground sources. When low-levels of flux are spuriously found in the photometry blue-ward of the 912Å break the photometric redshifts find a low-redshift solution. However, we also found cases at fainter flux levels where a low redshift solution was preferred because the photometric limits could not rule it out.

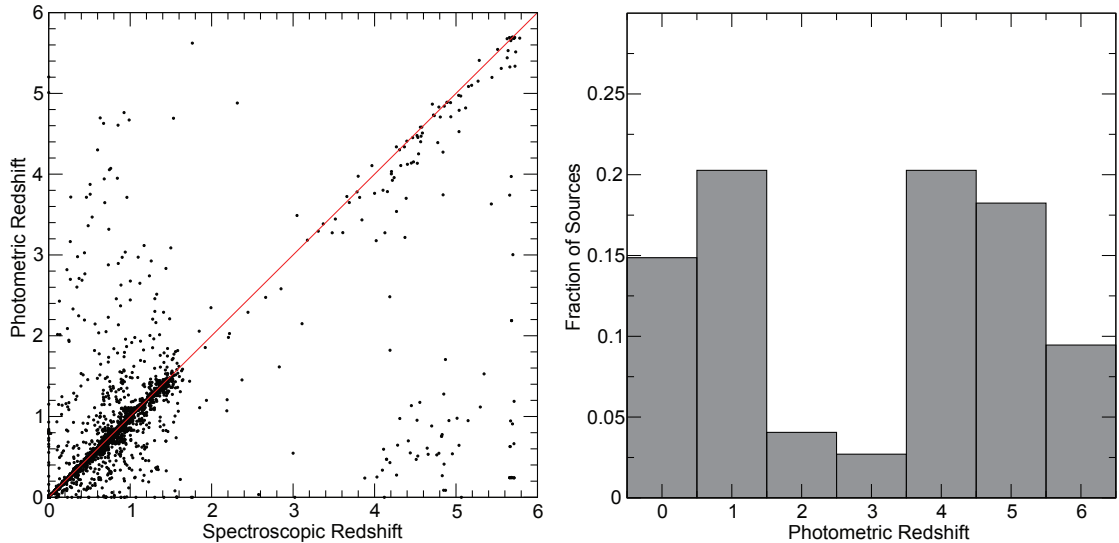

**Supplementary Figure 1:** A comparison of the photometric[30] and spectroscopic[31,32] redshifts for faint objects in the COSMOS field is shown on the left. In addition to the previously published results, this plot includes 167 objects at z>3 and >2200 objects fainter than I>22.5. In the right panel the photometric redshift distribution for 148 objects with spectroscopic redshifts at z>4 is shown. Note that 39.2% of objects are placed at z<3.

Rather than suffer the limitations of the photometric redshifts we chose to use a colour selection to identify potential proto-cluster members. The colour selection focuses on the region of the spectral energy distribution with maximum information. Furthermore, by limiting the number of filters used we avoid smoothing the images to the resolution of the worst seeing images and thereby maximizing the signal-to-noise of the photometry while minimizing contaminating light from foreground sources.

Photometry was measured using a procedure similar to that of the general COSMOS catalog[21]. However, rather than matching the point spread function (PSF) to 3" diameter apertures, the PSF homogenized images and a 1" and 2" diameter aperture was used. A statistical aperture correction was then applied to each band to bring the aperture magnitudes into average agreement with the 3" PSF matched aperture photometry from the COSMOS catalogue[21].

A simulated photometric catalogue was used to define a z~5.3 selection criteria. The simulation assumed the measured evolution of the UV luminosity function between 0<z<3 and no evolution at z>3. The distribution of galaxy types and obscuration was matched to the comprehensive spectroscopic sample in GOODS-N. However, the selection function is broadly insensitive to the exact assumptions made in the model, with decreased ages, decreased dust obscuration, and evolution at z>3 all narrowing the redshift distribution.

Using this simulation, we defined our colour criteria as no detection in a 1" diameter aperture at >3$\sigma$ significance in the u, B, g, and V bands along with colours of r-i>1 and IA738-i > 0.3. This selects a combination of z=5.36±0.2 galaxies, galactic M, L, T stars and faint (low mass) z~0.85 galaxies with old stellar populations and/or large obscuration.

Stars are easily removed because of their very red z-J colours but blue J-H colours. Low mass galaxies are removed using the spectral bump at 1.6$\mu$m if they are sufficiently bright in the J,H,K and IRAC bands. However, for objects fainter than H>23.5 ( i~25.5) (40% of our sample) we relied on the IRAC colours alone.

The uncertainty in the selection method has little impact on the conclusions. The objects with confirmed redshifts contain 65% of the proto-cluster mass, still within the range expected for a proto-cluster. Furthermore, the density of spectroscopically confirmed sources is significantly higher than that expected. Excluding objects without firm J,H,K detections only reduces the mass by 21% since these fainter objects tend to be the least massive.

## 4. Estimating the Magnitude and Significance of the Cluster

The significance of this cluster can be estimated from observational data[33,34] and from cosmological simulations[7]. Using our intermediate band colour selection to find galaxies at z=5.36±0.2 we find 8 objects in 1 square arc minute and 11 objects within a 2 Mpc (0.93 arc minute) radius of COSMOS AzTEC-3 brighter than I<26 ($L_*$ at z=5.3)[19]. These objects also meet the V and r dropout Lyman-Break criteria, but fall on the high-redshift end of these selections, so we consider V, r, and I band dropouts as comparison samples.

A conservative estimate of the expected surface density can be derived using the Great Observatories Origins Deep Survey (GOODS) data[19]. In these data a surface density of 0.75+/-0.04 galaxies per square arc minute brighter than $z_{850}$<26.0 is found in the combined V-band and I-band dropout samples. These samples cover a redshift range of 4.5<z<6.5, a factor of five more volume than our selection. Furthermore, objects are brighter in the $z_{850}$ filter than the I band, so the surface density is an over-estimate for an I band selected sample. However, even with this conservative estimate the region around COSMOS AzTEC-3 is over-dense by a factor of 11 in the 1 square arc minute region, and a factor of 5 within the 2 Mpc radius.

A better estimate of the significance can be obtained from the 4 square degree Canada-France-Hawaii Legacy Deep Survey (CFHT-LS Deep) data[20]. These data find a surface density of 0.352+/-0.006 r-dropout (4.5<z<5.2) galaxies per square arc minute brighter than z'<26 and a 2-point correlation function described by a clustering length ($r_o$) of 5.53 $h^{-1}$ Mpc and a power-law index of $\gamma=2.13$. Using the correlation function we estimate a variance of 1.06 galaxies in a 2 Mpc radius circle over the redshift interval probed by the r-band dropouts[20]. Using the mean density and variance, the cluster is 11.5 times over-dense and has a significance of 9.5$\sigma$.

The significance was also estimated using a simulated 2 square degree galaxy catalogue based on the Millennium simulation[35]. From the possible semi-analytics in this simulation we chose the WMAP 3yr cosmology, Galaxy formation Model C, and Dust Model 2 which most closely match the observed properties of z>3 galaxies. Galaxies were selected form the simulated catalogue to match the redshift distribution of the observed samples[20], and a magnitude limit of I<26.1 was adopted to match the observed object surface densities[20]. For the r-dropout selection function and a 2 Mpc radius cell we find a mean galaxy density of 0.94 and a variance of 1.14 galaxies per cell, meaning the observed cluster is an 11.7 times over-density with a significance of 8.8$\sigma$.

Alternatively, using the intermediate band selection function of z=5.36±0.2 we find a mean galaxy density of 0.14 and a variance of 0.39 galaxies per cell, a factor of 78.6 over-density, formally significant at 27.8$\sigma$. However, if we scale the r-band dropout surface density to the volume of the intermediate band selection and assume the same clustering properties, we find the significance is only 14.4$\sigma$.

Finally, if we consider only the three spectroscopically confirmed galaxies, a redshift interval of 5.3±0.005, and the measured r-band dropout clustering and space density we expect 0.024 galaxies in the 2Mpc cell with a variance of 0.19, implying a factor of 125 over-density significant at 16.7$\sigma$.

In summary, we find this region is over-dense by a factor of >11 at the >9$\sigma$ level, with the spectroscopic sub-sample suggesting an even higher significance.

**5. Estimating Stellar Masses**

To determine stellar masses models[24] were fit to the data red-ward of the Lyman-alpha forest break at rest frame 1216Å. These data included the IA827, NB816, z+, J, H, and Ks band photometry from the COSMOS catalog[21], and the IRAC 3.6μm and IRAC 4.5μm band photometry presented here. In the case of the extreme starburst the IRAC 5.8μm and 8.0μm detections[29] were also used. The properties of the proto-cluster members within 2 Mpc of the starburst are given in Supplementary Table 1. The optical ID numbers and positions are those from the COSMOS catalog[21], while the 3.6μm and 4.5μm fluxes are derived from the new Spitzer data described in supplementary section

1. The ages, dust obscuration (Av), and Masses are determined from our model fitting and are accurate to 3 Myr, 0.2 magnitudes and 0.3 dex respectively, however these errors are highly correlated and subject to systematic uncertainties in the underlying stellar population models.

**Supplementary Table 1: Model fitting results**

| Optical ID | RA J2000 | DEC J2000 | Age (Myr) | Av (Mag) | Stellar Mass (log10[$M_\odot$]) | 3.6µm flux (nJy) | 4.5µm flux (nJy) |
|---|---|---|---|---|---|---|---|
| 1445777 | 150.09148 | 2.602377 | 3 | 1.4 | 9.1 | 336±46 | 256±64 |
| 1445840 | 150.0937 | 2.600853 | 3 | 0.25 | 7.8 | <38 | <44 |
| 1446425 | 150.0839 | 2.598316 | 2.5 | 0.25 | 7.9 | <38 | <44 |
| 1446767 | 150.08328 | 2.595851 | 4.5 | 1.5 | 9.3 | 344±46 | 431±64 |
| 1447523 | 150.08573 | 2.589246 | 3 | 0.85 | 8.8 | 348±52 | 145±66 |
| 1447524* | 150.08958 | 2.575765 | 8 | 0.5 | 9.5 | 1372±48 | 2039±66 |
| 1447526 | 150.08654 | 2.586426 | 10 | 0.4 | 9.4 | 563±46 | 717±46 |
| 1447531*+ | 150.08625 | 2.58933 | 10 | 0.8 | 10.0 | 1419±48 | 2159±68 |
| 1448834 | 150.08309 | 2.588966 | 9 | 1.0 | 9.2 | 169±46 | 407±64 |
| 1449472 | 150.07305 | 2.582372 | 9 | 0.6 | 9.3 | 372±46 | 708±64 |

*Spectroscopically confirmed
+Starburst

## 6. Estimating The Infrared Luminosity of COSMOS AzTEC-3

The infrared luminosity was estimated empirically from the mm, radio limits, and CO line fluxes. The 1.1mm continuum measurement yields Far Infrared (60-120µm) luminosities ($L_{FIR}$) between 0.9-2 x$10^{13}$ $L_\odot$ assuming sub-mm slopes of 3.5 and 3 respectively. The actual spectral slope is constrained to be >3 at long wavelengths by the 870µm, 1.1mm and 2mm measurements[14,15] along with the 2.7, 3.2, and 8.2mm continuum limits obtained with the CO measurements[8], suggesting the source is cold and of lower intrinsic luminosity.

Evidence for cold dust is also supported by the flux ratios of the 870µm to 3.3mm data. The 870µm fluxes yield $L_{FIR}$ values that are 30% lower than those obtained with the 1.1mm flux alone and the 1.1mm flux yields a $L_{FIR}$ 80% lower than the 2mm flux alone. The 870µm data are near the dust emission peak at this redshift, and hence are likely to under-estimate $L_{FIR}$ if the dust temperature is low. Furthermore, the 1.1-3.3mm $L_{FIR}$ estimates are only consistent if the sub-mm spectral slope is >3.5.

Using the $L_{FIR}$ to radio correlation[16,36] with q=2.35 and a spectral slope of -0.8, we estimate $L_{FIR}$ <1.5 x$10^{13}$ $L_\odot$. Instead, using the L'CO- $L_{FIR}$ relation[37] we find $L_{FIR}$=1.3 x$10^{13}$ $L_\odot$.

Fitting models to the 24µm through radio photometry yields similar results given in Supplementary Table 2 and shown in Supplementary Figure 2. We note that the scatter in model estimates of $L_{IR}$ is much greater than the estimates of $L_{FIR}$. This reflects differing assumptions in the models since the data does not constrain rest frame wavelengths shorter than 140µm.

Combing these estimates we find $L_{FIR}$=1.7±0.8x$10^{13}$ $L_\odot$.

**Supplementary Table 2: Infrared model fitting results**

| Template | Model Name | $\chi^2$ | $L_{FIR}$ (60-120µm) ($L_\odot$) | $L_{IR}$(8-1000µm) ($L_\odot$) |
| --- | --- | --- | --- | --- |
| Chary & Elbaz[38] | $L_{IR}$=$10^{12.15}$ | 2.4 | 2.3x$10^{13}$ | 6.8x$10^{13}$ |
| Dale & Helou[39] | 1.1875 | 2.2 | 3.2 x$10^{13}$ | 1.1x$10^{14}$ |
| Lagache Hot[40] | $L_{IR}$=$10^9$ | 11.5 | 1.6 x$10^{13}$ | 3.4x$10^{13}$ |
| Lagache Cold[40] | $L_{IR}$=$10^{9.2}$ | 6.9 | 6.9 x$10^{12}$ | 2.2x$10^{13}$ |

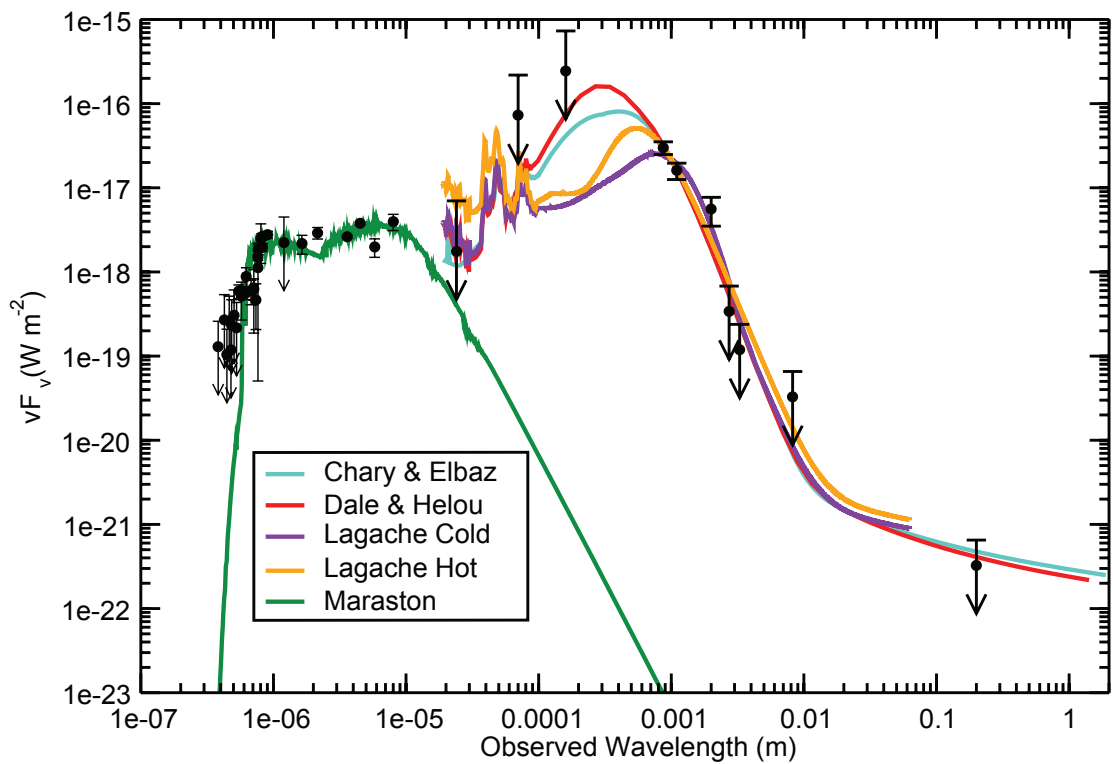

**Supplementary Figure 2:** The spectral energy distribution of COSMOS AzTEC-3 is shown along with model fits[24,38,39,40]. Error bars are 1σ including standard measurement error and systematic flux calibration error. The optical, near and mid-infrared, and radio data are taken from the COSMOS survey[16,21,29], with newly acquired warm IRAC data used at 3.6 and 4.5μm. The 0.87 and 1.1mm fluxes are taken from the sub-mm imaging survey[14,15], wile the 2mm data point presented here was acquired with the GISMO camera on the IRAM 30m telescope. The 2.7, 3.3, and 8.2mm flux limits are from the CO 6-5, 5-4, and 2-1 measurements[8]. The best-fit stellar mass is $1.0\pm0.3 \times 10^{10}$ solar masses with $A_v=0.8$ and an age of 10Myr. The best fit Far-Infrared (60-120μm) luminosity ranges from 0.9-3.2 $\times 10^{13}$ solar luminosities depending on the model assumptions. However, the 3mm upper limits, the radio flux limit, and the CO luminosity favor lower luminosities.